\documentclass[hyper]{JHEP} 

\usepackage{epsfig}

\newcommand\fverb{\setbox\pippobox=\hbox\bgroup\verb}

\newcommand\fverbdo{\egroup\medskip\noindent%

            \fbox{\unhbox\pippobox}\ }

\newcommand\fverbit{\egroup\item[\fbox{\unhbox\pippobox}]}

\newbox\pippobox

\title{Dynamics of  Probe Brane in the
Background of Intersecting Fivebranes}

\author{J. Kluso\v{n}
 \footnote{On leave from Masaryk University, Brno}\\
Dipartimento di Fisica,\\
Universita' di Roma \& Sezione di Roma 2, Tor Vergata \\
Via della, Ricerca, Scientifica, 1 00133  Roma   ITALY\\
E-mail:
\email{Josef.Kluson@roma2.infn.it}}

\preprint{ROM2F/2006/03 \\
\hepth{0601229}}

 \abstract{This paper is devoted to the study of the
  dynamics
 of the Dp-branes, F-strings and
M-branes in the  background of the system
 of two stacks of fivebranes in type IIA, IIB
and  M theory
that intersect on the line.} 


\def\bz{\mathbf{z}}
\def\by{\mathbf{y}}

\def\bA{\mathbf{A}}

\def\bAi{\left(\mathbf{A}^{-1}\right)}

\def\mL{\mathcal{L}}

\def\tR{S}
\begin{document}
\section{Introduction and Summary}\label{first}
The subject of intersecting branes
in string theory 
is very reach and it has been
studied  for a long time
\footnote{For review, see 
\cite{Gauntlett:2003di,Smith:2002wn,
Gauntlett:1997cv}.}.
Recently a new interesting
approach to this research
has been presented in the
work   \cite{Itzhaki:2005tu} where the
system consisting of two stacks of the
fivebranes in type IIB theory that intersect
on $R^{1,1}$ was studied from different points
of view. It was found, for example, that
\begin{itemize}
\item From the holography arguments one
could expect that the dynamics at the $1+1$
intersection of the two sets of fivebranes
should be holographically related to a
$2+1$ dimensional bulk theory, where the extra
dimension being the radial direction away
from the intersection. On the other hand
the bulk description contains two 
radial directions away from each set
of fivebranes and hence the bulk theory
is $3+1$ dimensional. Consequently the
corresponding boundary theory is $2+1$
dimensional.
\item Very interesting observation that
was given
in \cite{Itzhaki:2005tu} is  near horizon
symmetry enhancement 
\footnote{This observation
was also given from different
point of view in recent interesting paper 
\cite{Lin:2005nh}.}. Naivelly, looking
on the theory at the intersection of
the fivebranes one can deduce that this
theory should be invariant under $1+1$
dimensional Poincare symmetry $ISO(1,1)$.
However, the near horizon geometry
descibes a $2+1$ dimensional theory
with $ISO(2,1)$. 
\end{itemize}

Unfortunatelly there is not enough
place to review 
all results that were derived in
 \cite{Itzhaki:2005tu} and we recoomend
this paper for further reading.

In 
our previous papers 
\cite{Kluson:2005eb,Kluson:2005qq}
we have studied  
these  interesting properties
of the I-brane geometry \footnote{
In what follows we use the name
"I-brane" for the configuration of two 
stacks of fivebranes
intersecting on a line.}
from the point of D1-brane probe. We have
shown that the ehnancemment of the
near horiozon geometry 
has clear impact on the 
worldvolume theory of  the D1-brane probe.
In particular, we have
shown that the dynamics of
the 
D1-brane in the near horizon geometry
of I-brane can be interpreted
as the motion of the 
the probe D1-brane in  the
the near horizon geometry that
is sourced by the 
 object
that is extended in two spatial dimensions.
This result clearly demonstrates
an enhancemment of the symmetry 
studied from the I-brane worldvolume
theory in \cite{Itzhaki:2005tu}.

Our study of D1-brane probe in 
the near horizon geometry of I-brane
was based on two different
 approaches. In the fist
one we have shown that the D1-brane theory
posses additional scaling like symmetry
and we have determined the explicit
form of its generator. Using this
conserved charge and using also 
an existence of the conserved energy
we were able to solve the equation of
motion of the probe D1-brane explicitly.
The second way how to study the dynamics
of probe D1-brane was based on the 
transformations given in 
\cite{Itzhaki:2005tu}. Performing
these transformations on the worldvolume
of D1-brane we were able to map D1-brane
action to the  form where the enhancemment of
the symmetry was manifest. We have also
shown that these two approaches gave
the same picture of D1-brane dynamics in
near horizon region of I-brane background.

In this paper we will continue the 
study of the dynamics of the probe brane
in the background consisting of two stacks
of fivebranes intersecting on the line.
For simplicity we restrict ourselves
to the case of time dependent modes
on the worldvolume of probe only. 
We will  study the properties
of these probes in the near horizon region
of various geometries using the approach
similiar to the one that was given in 
\cite{Itzhaki:2005tu}. 
Namely, we will try to find transformation
that maps the probe action in given
near horizon background to the action
where an enhancemment of the symmetry
is manifest. Clearly we can
in principle proceed in the same way
as in \cite{Kluson:2005eb,Kluson:2005qq}
 and define
the second conserved charge corresponding
to the new symmetry that has the form
of scaling like symmetry 
in the original action. Since these
two approaches are equivalent we restrict
ourselves to the approach based on 
\cite{Itzhaki:2005tu} for its   elegance
and simplicity. 

The plan of this paper is as follows. 
 In the next section we will
 discuss a D1-brane
probe in the background of two stack
of D5-branes intersecting on line in
type IIB theory. We will see that
in the near horizon region the enhancemment
symmetry is again manifest. 
We will also see that the dynamics
of probe D1-brane with zero electric
flux is trivial as a consequence of
the fact that the configuration of
D5-branes and D1-brane is supersymmetric.

In section (\ref{third}) we will
study the probe F-string in the
background of NS5-branes intersecting
on a line in type IIA theory. 
 We will see that again
the dynamics of F-string
in given background is very simple
 as a consequence
of the fact that the configuration
of F-string that is extended along
directions  paralell with the
worldvolumes of
NS5-branes is supersymmetric. 
Then we will address the question
of Dp-brane as a probe in I-brane
configuration in type IIA theory. 
Since there is not any stable D1-brane,
all worldvolume theories  of 
Dp-branes with ($p\neq 0$) will
explicitly depend on the
wordlvolume spatial coordinate
as a consequence of the fact that
the Poincare symetry is preserved in the
$R^{1,1}$ subspace only. This
observation implies that it is not
possible to find configurations
that depend on time only.
We will also discuss the question
of D0-brane as a probe. We 
find that its dynamics is completely
equivalent to the dynamics of D1-brane
in the background of I-brane 
of type IIB theory. We will
argue that this is a consequence
of the T-duality that relates
type IIA to type IIB theory vice
versa. 
Finally in section (\ref{fourth})
we will consider a M2-brane probe
in the background of two stacks
of M5-branes intersecting on line
that are however delocalised in
the common transverse dimension.
We will again see that its dynamics
is equivalent to the dynamics
of probe F-string in I-brane
background in type IIA theory.

Let us now  outline our results.
 We mean that
they clearly demonstrate that
the approach of the study of the
intersecting geometries based on
D-brane probe in the near horizon
region is very powerful. In fact, 
the importance of the Dp-brane
as a probe has been known for long
time \cite{Tseytlin:1996hi,Bozhilov:2002sj}
\footnote{For review, see for example
\cite{Johnson:2000ch}.}
 however as far as we know
the study of the I-brane geometry using
Dp-brane probe has not been performed
so far. For that reason we hope
that our modest contribution could
be useful for the further study of the
properties of I-branes. 

\section{D1-brane in the D5-D5' background}
\label{second}
We start our discussion with the
 analysis of the dynamics of 
probe D1-brane in the background
of two stacks of D5-brane
that intersect on the line.
More precisely, we
have $k_1$ D5-branes extended in
$(0,1,2,3,4,5)$ direction and the set
of $k_2$ D5-branes extended in
$(0,1,6,7,8,9)$ directions. Let us
define
\begin{eqnarray}
\by=(x^2,x^3,x^4,x^5) \ , \nonumber \\
\bz=(x^6,x^7,x^8,x^9) \ .
\nonumber \\
\end{eqnarray}
We have $k_1$ D5-branes localized
at the points $\bz_n \ n=1,\dots,k_1$
and $k_2$ D5-branes localized
at the points $\by_a \ , a=1\dots,k_2$.
Every pairs of fivebranes from
different sets intersect at different
point $(\by_a,\bz_n)$.
The background geometry corresponding to this
configuration has the form 
\begin{eqnarray}\label{dd5bac}
d^2s=(H_1H_2)^{-1/2}
(-dt^2+(dx_1)^2)+H_1^{-1/2}
H_2^{1/2}\delta_{\alpha\beta}dx^\alpha
dx^\beta+H^{1/2}_1
H^{-1/2}_2\delta_{pq}dx^pdx^q \  
\nonumber \\
\end{eqnarray}
together with nontrivial
dilaton  
\begin{equation}
e^{2\Phi}=(g_1g_2)^2H^{-1}_1H^{-1}_2 \ ,
\nonumber \\
\end{equation}
where $\Phi=\Phi_1(\bz)+\Phi_2(\by)$ and
\begin{equation}
e^{2(\Phi_1-\Phi_1(\infty))}=
\frac{1}{g_1^2}e^{2\Phi_1}=H^{-1}_1 \ , 
e^{2(\Phi_2-\Phi_2(\infty))}=
\frac{1}{g_2^2}e^{2\Phi_2}=
H^{-1}_2 \ . 
\end{equation}

Finally, we also have 
RR three form
\begin{eqnarray}
 H_{\alpha\beta\gamma}=
-\epsilon_{\alpha\beta\gamma\delta}
\partial^\delta \Phi_2 \ ,
\alpha,\beta,\gamma,\delta=
2,3,4,5 \ , \nonumber \\
H_{pqr}=-\epsilon_{pqrs}
\partial^s\Phi_1 \ ,
p,q,r,s=6,7,8,9 \ , \nonumber \\
\end{eqnarray}
 where
\begin{eqnarray}\label{H12g}
H_1=1+
\sum_{n=1}^{k_1}
\frac{l_s^2}{|\bz-\bz_n|^2} \ ,
\nonumber \\
H_2=1+\sum_{a=1}^{k_2}
\frac{l_s^2}{|\by-\by_a|^2} \ .
\nonumber \\
\end{eqnarray}
Our goal is to study
properties of this background
from the point of view of D1-brane
probe when $\bz_n=\by_a=0$.
In this case (\ref{H12g})
takes the form
\begin{equation}
H_1=1+\frac{\lambda_1}{|\bz|^2} \ , 
H_2=1+\frac{\lambda_2}{|\by|^2} \ , 
\lambda_1=N_1l_s^2 \ , \lambda_2=N_2l_s^2
\ . 
\end{equation}
Let us now consider 
the probe  D1-brane. Recall
that the dynamics of the D1-brane
is governed by Dirac-Born-Infeld (DBI) action
\begin{equation}\label{d1g}
S=-\tau_1\int d^2\xi
e^{-\Phi}\sqrt{-\det\bA} \ , 
\end{equation}
where $\tau_1$ is D1-brane tension,
$\xi^\mu \ , \mu=0,1$ are worldvolume
coordinates and where the matrix 
$\bA_{\mu\nu}$ is equal to
\begin{equation}
\bA_{\mu\nu}=g_{MN}\partial_\mu
X^M\partial_\nu X^N+\partial_\mu
A_\nu-\partial_\mu A_\nu \ ,
\end{equation}
where $A_\mu$ is worldvolume
gauge fields 
\footnote{We work in units
$2\pi \alpha'=1$.}
and $X^M\ , M=0,\dots,9$ paremetrise
the position of D1-brane in targed
spacetime. 

Let us now presume that
D1-brane is stretched in $x^0,x^1$
directions.  Then
it is natural to choose the
 static gauge
where $x^0=X^0 \ , x^1=X^1$
and hence in the background
(\ref{dd5bac}) the action
(\ref{d1g}) takes the form
\begin{equation}
S=-\tau_1\int
d^2x e^{-\Phi}\sqrt{-\det\bA} \ ,
\end{equation}
where
\begin{equation}
e^{-\Phi}=\frac{H_1^{1/2}
H_2^{1/2}}{g_1g_2}
 \ , 
\bA_{\mu\nu}=
g_{\mu\nu}+g_{pq}\partial_\mu
Z^p\partial_\nu Z^q+
g_{\alpha\beta}
\partial_\mu Y^\alpha\partial_\nu
Y^\beta+\partial_\mu A_\nu-
\partial_\nu A_\mu \ .
\end{equation} 
Note  that there is also
WZ term in the form
\begin{equation}\label{swc}
S_{WZ}=
\frac{\tau_1}{2}\int d^2x
\epsilon^{\mu\nu}\left(
C_{pq}\partial_\mu Z^p
\partial_\nu Z^q+
C_{\alpha\beta}\partial_\mu
Y^\alpha \partial_\nu Y^\beta
\right)
\end{equation}
that expresses the coupling
of the D1-brane to the Ramond-Ramond
two form $C$.

Now we restrict ourselves
on the study of the 
dynamics of the probe D1-brane
in the near horizon region of
the I-brane background. As we
 claimed in the introduction,
we can  proceed in two ways. The
first one that was previously
introduced in  \cite{Kluson:2005eb}
is based on an existence of an additional
symmetry that
emerges  in the near horizon background
for pure time dependent motion of
the probe D1-brane. Then using corresponding
conserved charge and also using 
conserved energy of the probe we 
can solve its equations of motion explicitly.
The second approach is based on 
the beautiful observation 
\cite{Itzhaki:2005tu,Lin:2005nh}
that in the near horizon region of I-brane
an enhancemment of the symmetry of the
transverse space emerges. Since it seems to us
that the second approach gives more physical
insaight into  the I-brane physics  
we proceed in the similar way here as well.
Namely,  we 
will try to find the transformation in the
D1-brane DBI-action in the near horizon
region of I-brane that maps this action
to the equivalent one where the ehnancemment
of the symmetry will be manifest.  As
the final remark note that for homogeneous
modes the WZ term (\ref{swc})
does not contribute 
to the equations of motion. 

Now thanks to the manifest rotation
symmetry $SO(4)$ in the subspaces
spanned by coordinates $\bz=(z^6,z^7,z^8,z^9)$ and
$\by=(y^2,y^3,y^4,y^5)$ we will
reduce the problem to the study of the
motion of two dimensional subspaces, namely
we will presume that only following wordldvolume
modes are excited
\begin{equation}
z^6=R \cos \alpha \ ,
z^7=R \sin \alpha \ ,
\end{equation}
and
\begin{equation}
y^2=\tR \cos\beta \ ,
y^3=\tR \sin\beta \ .
\end{equation}
We also presume that
we have nonzero $\partial_0 A_1\equiv
\dot{A}$ while we work in gauge
$A_0=0$. 
Then the matrix $\bA$ takes the form
\begin{equation}
\bA=\left(
\begin{array}{cc}
\frac{RS}{\sqrt{\lambda_1
\lambda_2}}
\left(-1+\frac{\lambda_2}{S^2}
(\dot{S}^2+S^2\dot{\beta}^2)+
\frac{\lambda_1}{R^2}(
\dot{R}^2+R^2\dot{\alpha}^2)\right)
& \frac{\sqrt{\lambda_1
\lambda_2}}{RS}\dot{A} \\
-\frac{\sqrt{\lambda_1
\lambda_2}}{RS}\dot{A}
& \frac{RS}{\sqrt{\lambda_1
\lambda_2}} \\ \end{array}
\right)
\end{equation}
where $x^0\equiv t \ , 
\dot{X}=\partial_0(X)$. 
Since the worldvolume theory
does not depend on $x^1$ it 
is convenient to work with
the action density $s\equiv \frac{S}{V}$
where $V$ is the volume of
the worldvolume spatial section 
that in case of D1-brane is simply
length $L$ of D1-brane. 
Using this definition we 
get
\begin{equation}\label{sii}
s\equiv \frac{S}{V}=
-\int dt 
\frac{1}{g_1g_2}
\sqrt{1-\frac{\lambda_2}{S^2}
(\dot{S}^2+S^2\dot{\beta}^2)-
\frac{\lambda_1}{R^2}(
\dot{R}^2+R^2\dot{\alpha}^2)
-\frac{\lambda_1^2\lambda_2^2}
{R^4S^4}\dot{A}^2} \ .
\end{equation}
Following the work of \cite{Itzhaki:2005tu}
we now  introduce two modes
\begin{equation}
R=e^{\frac{2}{k_1}\phi_1} \ ,
S=e^{\frac{2}{k_2}\phi_2} \ 
\end{equation}
and perform the linear
transformation  in
the form
\begin{equation}\label{phi12}
\frac{1}{k_1}\phi_1+\frac{1}{k_2}\phi_2=A\phi \ ,
\phi_1-\phi_2=B x^2 \ . 
\end{equation}
Coeficients $k_1,k_2,A,B$ are fixed
by requiraments that the
kinetic terms for $\phi,x_2$ take cannonical
form  $\dot{\phi}^2+\dot{x}_2^2$.
This condition can be obeyed with
following values of coeficints 
\begin{eqnarray}
A=\frac{1}{2}\sqrt{\frac{1}{\lambda}} \ ,
 \frac{1}{\lambda}=\frac{1}{\lambda_1}+
\frac{1}{\lambda_2} \ , 
B=\frac{\lambda_1}{2} \ ,
\nonumber \\
k_2=\lambda_2 \ , k_1=\lambda_1 \ . 
\nonumber \\ 
\end{eqnarray}
After performing this
transformation
the action 
(\ref{sii})  takes the form
\begin{equation}\label{si}
s=-\int dt\frac{\tau_1}{g_1g_2}
\sqrt{1-
\dot{\phi}^2-
\dot{x}_2^2
-\lambda_1\dot{\alpha}^2
-\lambda_2\dot{\beta}^2-
(\lambda_1\lambda_2)^2e^{-\frac{1}{\sqrt{\lambda}}
\phi}\dot{A}^2}
 \ .
\end{equation} 
From (\ref{si}) we immediatelly
see that the enhancemment of
the near horizon geometry 
takes place even in the D5-D5'
background. In fact, from the
point of view of D1-brane probe
the direction $x_2$ is equivalent
to $x^0,x^1$ directions. In other
words the action (\ref{si}) has the form of
the action for a  D1-brane  
that moves in  in the near horizon
region of some  $2+1$ dimensional
object.

Since I-brane background is
time independent the
 energy of D1-brane 
is conserved and hence it is
natural to study its dynamics 
using the Hamiltonian formalism.
For this reason  we will now be more
general and consider the
Lagragian in the form
\begin{equation}\label{lagG}
\mL=-\sqrt{V-\sum_i (f_i
(\partial_0{\Phi}^i)^2
+B_i\partial_0\Phi^i)}\equiv -\triangle \ ,
\end{equation}
where $V$ is a  potential
for dynamical  fields $\Phi^i$ and
where $f_i$ and $B_i$ are functions
of $\Phi^j$.  
The conjugate momentum $P_i$ to $
\Phi^i$ takes the form
\begin{equation}\label{Pg}
P_i=\frac{2f_i\partial_0\Phi^i+B_i}{2\triangle}
\ 
\end{equation}
and hence
\begin{equation}\label{ophi}
 \partial_0\Phi^i=\frac{1}{2f_i}
\left(2P_i\triangle-B_i\right) \ . 
\end{equation}
Then using (\ref{Pg}) and
(\ref{ophi}) we get the  Hamiltonian
in the form
\begin{equation}
H=\sum_iP_i\partial_0\Phi^i-\mL=
\frac{2V-\sum_i B_i\partial_0\Phi^i}{2\triangle}
=\frac{2V+\sum\frac{B_i^2}{2f_i}}{2\triangle}
-\sum_i
\frac{B_i}{2f_i} P_i \ . 
\end{equation}
If we express the Hamiltonian
using the cannoncial variables
$\Phi^i,P_i$ we get
\begin{eqnarray}\label{Hgen}
H=\sqrt{\left(V+\sum_i\frac{B_i^2}{4f_i}\right)
\left(1+\sum_i\frac{P^2_i}{f_i}\right)}
-\sum_i\frac{B_iP_i}{2f_i} \ . 
\nonumber \\
\end{eqnarray}
In case of the  action (\ref{si}) we have
\begin{equation}
V=\frac{\tau_1^2}{(g_1g_2)^2} \ ,
f_{\phi}=V \ ,
f_{x^2}=V \ ,
f_{\alpha}=\lambda_1 V \ ,
f_{\beta}=\lambda_2 V \ ,
f_{A}=V(\lambda_1\lambda_2)^2
e^{-\frac{1}{\sqrt{\lambda}}\phi}
\end{equation}
and $B_i=0$ so that  we obtain from
(\ref{Hgen}) the Hamiltonian
in the form
\begin{eqnarray}\label{h}
H=\sqrt{\frac{\tau_1^2}{(g_1g_2)^2}
+P_\phi^2+P_{x^2}^2+\frac{1}{\lambda_1}
P_\alpha^2+\frac{1}{\lambda_2}P_\beta^2
+\frac{e^{\frac{1}{\sqrt{\lambda}}
\phi}}{(\lambda_1\lambda_2)^2}
\pi^2} \ ,  \nonumber \\
\end{eqnarray}
where $\pi=P_A$. 
Now the cannonical equations of motion for
$P_\alpha,P_\beta,P_2$ and $\pi$ 
can be easily derived from (\ref{h})
and we get
\begin{equation}
\dot{P}_{\Phi^i}=-\frac{\delta H}{\delta \Phi^i}=0 \ ,
\dot{\pi}=-\frac{\delta H}{\delta A}=0 \ , 
\end{equation}
where $\Phi^i=\left(\alpha,\beta,
x^2\right)$. 
We see that the problem reduces
to the study of dynamics of $\phi$
that obeys the cannonical
equation of motion 
\begin{equation}\label{dotphi}
\dot{\phi}=\frac{\delta 
H}{\delta P_\phi}
=\frac{P_\phi}{E}
\end{equation}
using the fact that $H$ is conserved
and equal to the energy  $E$. 
To solve this equation we express 
$P_\phi$ from (\ref{h}) and we get
\begin{eqnarray}
P_\phi=\pm
\sqrt{E^2-\frac{\tau_1^2}{(g_1g_2)^2}-
P_{x^2}^2
-\frac{1}{\lambda_1}P_\alpha^2
-\frac{1}{\lambda_2}P_\beta^2
-\frac{e^{\frac{1}{\sqrt{\lambda}}
\phi}}{(\lambda_1\lambda_2)^2}
\pi^2}
\equiv
\nonumber \\
\sqrt{G^2-H^2e^{\frac{1}{\sqrt{\lambda}}
\phi}} \ , \nonumber \\
\end{eqnarray}
where
\begin{equation}
G^2=E^2-\frac{\tau_1^2}{(g_1g_2)^2}-P_{x^2}^2
-\frac{1}{\lambda_1}P_\alpha^2
-\frac{1}{\lambda_2}P_\beta^2 \ ,
H^2=\frac{\pi^2}{(\lambda_1\lambda_2)^2}
 \ .
\end{equation}
Then the equation (\ref{dotphi})
takes the form
\begin{equation}
\dot{\phi}=\pm \frac{1}{E}
\sqrt{G^2-H^2e^{\frac{1}{\sqrt{\lambda}}
\phi}} \ .
\end{equation}
First of all we see that for $\pi=0$  
the equation above has simple solution 
\begin{equation}
\phi=t\frac{\sqrt{
E^2-\frac{\tau_1^2}{(g_1g_2)^2}-P_{x^2}^2
-\frac{1}{\lambda_1}P_\alpha^2
-\frac{1}{\lambda_2}P_\beta^2}}{E}
+\phi_0
\end{equation}
that corresponds to the motion of 
free particle. 
In particular, for $P_\phi=0$ we get $\phi=\phi_0$ and
we obtain the solution corresponding to D1-brane
at the rest at the distance $\phi_0$ from the
line of 
intersection of D5-branes.
 This is satisfactory result since 
the configuration of  two stacks of
D5-branes intersecting on line  and D1-brane
with the worldvolume paralell with the
line of intersection is supersymmetric
and hence   D1-brane at rest
can be localised  at any distance from
the intersection. Of course this 
 is valid for any $R,S$ as follows from the fact that 
the original Lagrangian for D1-brane
in I-brane background does not contain any potential 
for $R,S$ when D1-brane does not move.
In other words,  it can be easily shown
that the solution of the equation of motion for
vanishing $\dot{R},\dot{S}$ exists for any $R,S$.

The situation is different for
$\pi\neq 0$. Note that $\pi$ measures the
number of fundamental strings
that form the bound state with
D1-brane. On the other hand it is well
known that the probe F-string
in given background breaks all
supersymmetries. The manifestation
of this fact is the  presence of
the potential in  Hamiltonian
(\ref{h}). In spite of this
fact we can solve the equation 
(\ref{dotphi}) and we get
\begin{equation}\label{sol}
e^{\frac{1}{\sqrt{\lambda}}
\phi}=\frac{G^2}{H^2}\frac{1}{
\sinh^2\frac{G}{4E\sqrt{\lambda}}t} \ , 
\end{equation}
where we have choosen
 the initial
condition that for $t=0$ D1-brane
reaches  its turning point $\dot{\phi}=0$.

The physical picture of the result
of (\ref{sol}) is standart:
 D1-brane leaves the worldvolume
of I-brane at $t=-\infty$ reaches
its turning point at $t=0$ and
goes back to I-brane for $t=\infty$.

In conclussion of this
section we  would
like to stress the main result 
determined here. Namely
we have  shown that
the background of two orthogonal
stacks of D5-branes posses
an ehnancemment of the symmetry
in the near horizon region exactly
in the same way as in the case
of two stacks of interescting
NS5-branes. 
\subsection{Dynamics of
D1-brane near the points
$R_t^2=\lambda_1 \ ,
S_t^2=\lambda_2$}
The analysis performed in 
\cite{Itzhaki:2005tu} 
suggests an existence of two
exception points $R_t$ and $S_t$ whose
values have been defined in the title
of this section. 
Namely, it was argued
in 
\cite{Itzhaki:2005tu} 
that free fermions that for
zero coupling  are 
supported on $R^{1,1}$
are for nonzero gauge coupling
displaced from  $R^{1,1}$. More preciselly,
fermions living on intresection
can be replaced by holomorphic
current algebra
\begin{equation}
SU(k_1)_{k_2}
\times SU(k_2)_{k_1}
\times U(1) \ .
\end{equation}
The carefoul analysis of the
low energy effective theory on I-brane
performed in  
\cite{Itzhaki:2005tu}
leads to following picture. 
At vanishing coupling $g_1=g_2=0$
$k_1k_2$ fermions are localised on the
at point $r=s=0$ as suggested by
their description in terms of open strings
stretched between D5-branes. The turning
on the coupling constant moves these
fermions from the origin. Different currents
that are formed from different fermions
move in different directions. In particular,
the $SU(k_1)_{k_2}$ move to 
$s \sim \sqrt{\lambda_2}$. For 
$SU(k_2)_{k_1}$ one moves to 
$r\sim \sqrt{\lambda_1}$. The $U(1)$ part
is supported at both points. 

Our question is whether this
interesting behaviour of the
low energy I-brane worldvolume
theory has some impact on the
dynamics of the D1-brane probe
in I-brane background. For that
reason we will  study the
dynamics of the D1-brane near the
points $R_t,S_t$ defined above.
Then it is convenient to  write
\begin{equation}
R=r+R_t \ , r\ll R_t \ ,
S=s+S_t \ , s\ll S_t \ 
\end{equation}
and hence 
\begin{eqnarray}
H_1=1+\frac{\lambda_1}{(\sqrt{\lambda_1}+r)^2}
\approx 2(1-\frac{r}{\sqrt{\lambda_1}})
\nonumber \\
H_2=1+\frac{\lambda_2}{(\sqrt{\lambda_2}+s)^2}
\approx 2(1-\frac{s}{\sqrt{\lambda_2}}) \ .
\nonumber \\
\end{eqnarray}
With this approximation
the DBI action for D1-brane
near the points $R_t,S_t$ 
takes the form 
\begin{equation}\label{sapprox}
s=-\frac{\tau_1}{g_1g_2}
\int dt
\sqrt{1-2\dot{s}^2-2\dot{r}^2
-2\lambda_1\dot{\alpha}^2-2
\lambda_2\dot{\beta}^2
-4\left(1-\frac{r}{\sqrt{\lambda_1}}
-\frac{s}{\sqrt{\lambda_2}}\right)} \ ,
\end{equation}
where we have considered
the terms at most linear in $r,s$ and
we have  restricted on homogeneous
modes only.
We again introduce two modes $x,\phi$ defined
as 
\begin{equation}
r=M^r_xx+M^r_\phi \phi \ , 
s=M^s_xx+M^s_\phi \phi \ , 
\end{equation}
where $M^i_j \ , i=r,s \ , j=x,\phi$ are
coeficients  that will be 
determined from following
requiraments: Firstly, we demand that
 the kinetic term $2\dot{r}^2+
2\dot{s}^2$ takes the
cannonical  form in new variables
$\phi$ and $x$.  
Secondly, we  demand that
the following expression
\begin{equation}
\frac{r}{\sqrt{\lambda_1}}+
\frac{s}{\sqrt{\lambda_2}}=
\left(\frac{M^r_x}{\sqrt{\lambda_1}}
+\frac{M^s_x}{\sqrt{\lambda_2}}\right)x^2
+\left(
\frac{M^r_\phi}{\sqrt{\lambda_1}}
+\frac{M^s_\phi}{\sqrt{\lambda_2}}\right)\phi
\end{equation}
depends on $\phi$ only. 
These conditions imply
\begin{eqnarray}
M^r_\phi=\frac{\sqrt{\lambda_2}}
{\sqrt{2}\sqrt{\lambda_1+\lambda_2}} \ , 
 M^s_\phi=\frac{\sqrt{\lambda_1}}
{\sqrt{2}\sqrt{\lambda_1+\lambda_2}}
\nonumber \\
\frac{M^r_\phi}{\sqrt{\lambda_1}}
+\frac{M^s_\phi}{\sqrt{\lambda_2}}
=\frac{1}{\sqrt{2\lambda}} \ ,
\frac{1}{\lambda}=
\frac{1}{\lambda_1}+\frac{1}{\lambda_2}
\nonumber \\
M^s_x=\frac{\sqrt{\lambda_2}}{\sqrt{2}\sqrt{\lambda_1+
\lambda_2}} \ ,
M^r_x=-\frac{\sqrt{\lambda_1}}
{\sqrt{2}\sqrt{\lambda_1+\lambda_2}} \ .
\nonumber \\  
\end{eqnarray}
Then  the action 
(\ref{sapprox})
takes the form
\begin{equation}\label{snb}
S=-\frac{\tau_1}{g_1g_2}
\int dt 
\sqrt{1-\dot{x}^2-\dot{\phi}^2-2\lambda_1\dot{\alpha}^2
-2\lambda_2\dot{\beta}^2
-4\left(1-\frac{\phi}{\sqrt{2\lambda}}
\right)\dot{A}^2} \ .
\end{equation}
Now we can easily study the
dynamics of the D1-brane near
the points $R_t,S_t$. 
We again introduce the Hamiltonian
following the genear discussion given
in previous subsection. 
Since for (\ref{snb}) we have
\begin{eqnarray}
V=\left(\frac{\tau_1}{g_1g_2}
\right)^2
 \ ,
B_i=0 \ ,
f_x=V \ , f_\phi=V \ ,
\nonumber \\
f_\alpha=2\lambda_1 V \ ,
f_\beta=2\lambda_2V \ ,
f_A=4V\left(1-\frac{\phi}
{\sqrt{2\lambda}}\right) 
\nonumber \\ \  
\end{eqnarray}
it is clear that the
 Hamiltonian (\ref{Hgen})
 is equal to
\begin{equation}\label{Hgennsr}
H=\sqrt{
\left(\frac{\tau_1}{g_1g_2}
\right)^2
+P^2_\phi+P^2_x+
\frac{P^2_\alpha}{2\lambda_1}
+\frac{P^2_\beta}{2\lambda_2}
+\pi^2\left(1-\frac{\phi}{
\sqrt{2\lambda}}\right)^{-1}}
\ .
\end{equation}
We again see that $P_x,P_\alpha,P_\beta$ and
$\pi$ are conserved so that
the problem reduces to the
study of the dynamics of the
mode $\phi$. Using (\ref{Hgennsr})
we get that it obeys the
equation of motion 
\begin{equation}\label{dotphirs}
\dot{\phi}=\frac{\delta H}
{\delta P_\phi}=\frac{P_\phi}
{\sqrt{(\dots)}}=
\frac{P_\phi}
{E} \ , 
\end{equation}
where $E$ is conserved energy.
Now from (\ref{Hgennsr})
 we express $P_\phi$ as
\begin{equation}
P_\phi=
\pm\sqrt{
E^2-\left(\frac{\tau_1}{g_1g_2}
\right)^2
-P^2_x-
\frac{P^2_\alpha}{2\lambda_1}
-\frac{P^2_\beta}{2\lambda_2}
-\pi^2\left(1-\frac{\phi}{
\sqrt{2\lambda}}\right)^{-1}}
\end{equation}
and hence (\ref{dotphirs})
implies 
\begin{equation}\label{dotphirs1}
\frac{d\phi}
{\sqrt{G^2-H^2\phi}}
=\pm dt \ , 
\end{equation}
where we have used
the fact that $\phi\ll \sqrt{\lambda}$.
Note also that 
$G^2$ and $H^2$ given
in (\ref{dotphirs1})  are defined
as  
\begin{eqnarray}
G^2=E^2-
P^2_x-
\frac{P^2_\alpha}{2\lambda_1}
-\frac{P^2_\beta}{2\lambda_2}
-\pi^2
-\left(\frac{\tau_1}{g_1g_2}
\right)^2
\ , \nonumber \\
H^2=\frac{1}{\sqrt{2\lambda}}
\pi^2 \ .
\nonumber \\
\end{eqnarray}
By trivial integration 
of (\ref{dotphirs1})
we get
\begin{equation}
\phi=\frac{G^2}{H^2}
-\frac{1}{H^2}
\left(\pm\frac{H^2}{2}t+C
\right)^2 \ , 
\end{equation}
Where $C$ is an integration constant. 
If we choose the initial condition that
for $t=0$ $\phi$ is equal to zero
we get $C=G$. Then the final
result takes the form
\begin{equation}
\phi=
\pm\frac{t}{2}-\frac{H^2}{4G^2}
t^2 \ .
\end{equation}
We  see that for $\pi=0$
we obtain the trajectory of free
particle which  is again a consequence of
the fact that configuration of
two stacks of D5-brane and D1-brane
as was defined above is supersymmetric.

We also see that
 the interesting properties of the
I-brane worldvolume theory
that were review above do
not have any impact on the
 dynamics of
D1-brane probe. This is
of course a  naturally result since the
supergravity background corresponds
to the vacuum state of the I-brane
worldvolume theory where  no
I-brane worldvolume modes  are excited.
For that reasion it would be certainly
nice to  try to generalise the approach
presented in \cite{Lin:2005nh}
to the case of I-brane theory as well.
Namely, we can presume that 
the condensation of the worldvolume
modes will lead to the deformation
of the background geometry. Then
it would be interesting to 
study the question how the motion
of D1-brane probe in such a background
depends on the condensation of the
I-brane worldvolume fields.

However it is still very interesting
that D1-brane near the points 
$R_t,S_t$ sees an enhancemment 
of the symmetry as well. It would
be also nice to underestand  
this observation  better.

\section{ F-string as a probe
of  I-brane Background in
type IIA theory}\label{third} 
In this
section we will study the dynamics of
the macroscopic string  in the
background of two stacks of NS5-branes
intresecting on line in 
 type IIA theory. More preciselly,
 we have configuration of 
 $k_1$ NS5-branes extended in
$(0,1,2,3,4,5)$ direction and the set
of $k_2$ NS5-branes extended in
$(0,1,6,7,8,9)$ directions. We 
again define 
\begin{eqnarray}
\by=(x^2,x^3,x^4,x^5) \ , \nonumber \\
\bz=(x^6,x^7,x^8,x^9) \ .
\nonumber \\
\end{eqnarray}
We have $k_1$ NS5-branes localized
at the points $\bz_n \ n=1,\dots,k_1$
and $k_2$ NS5-branes localized
at the points $\by_a \ , a=1\dots,k_2$.
The supergravity background corresponding
to this configuration takes
the form
\begin{eqnarray}
\Phi(\bz,\by)=\Phi_1(\bz)+
\Phi_2(\by) \ , \nonumber \\
g_{\mu\nu}=\eta_{\mu\nu} \ ,
\mu,\nu=0,1 \ , \nonumber \\
g_{\alpha\beta}=e^{2(\Phi_2-
\Phi_2(\infty))}\delta_{\alpha\beta} \
, H_{\alpha\beta\gamma}=
-\epsilon_{\alpha\beta\gamma\delta}
\partial^\delta \Phi_2 \ ,
\alpha,\beta,\gamma,\delta=
2,3,4,5\nonumber \\
g_{pq}=e^{2(\Phi_1-\Phi_1(\infty))}
\delta_{pq} \ ,
H_{pqr}=-\epsilon_{pqrs}
\partial^s\Phi_1 \ ,
p,q,r,s=6,7,8,9 \ , \nonumber \\
\end{eqnarray}
where $\Phi$ on the
first line means the dilaton
and where
\begin{eqnarray}
e^{2(\Phi_1-
\Phi_1(\infty))}=1+
\sum_{n=1}^{k_1}
\frac{l_s^2}{|\bz-\bz_n|^2}=H_1(\bz) \ ,
\nonumber \\
e^{2(\Phi_2-
\Phi_2(\infty))}=1+\sum_{a=1}^{k_2}
\frac{l_s^2}{|\by-\by_a|^2}=H_2(\by)  \ .
\nonumber \\
\end{eqnarray}
The dynamics of the fundamental macroscopic
 string
in this background is governed by 
Nambu-Gotto action 
\begin{equation}\label{stringfund}
S=-\tau_s\int d\tau d\sigma
\sqrt{-\det\bA}+S_{WZ} \ ,
\end{equation}
where $\sigma\in (-\infty,
\infty)$ and 
where $S_{WZ}$ expresses the coupling
of the string to the two form $B$.
Since we will be interested in the
dynamics of the string  with
the worldsheet alligned along
the worldvolume of  I-brane 
and with the time dependent worldsheet
fieds only the Wess-Zumino term 
 does not contribute to the
equation of motions. 
In (\ref{stringfund}) the matrix $\bA$ is 
\begin{equation}
\bA_{\mu\nu}=
 \partial_\mu X^M\partial_\nu X^N g_{MN}
  \ . 
\end{equation}
Then the equations of motion that 
arrise from (\ref{stringfund}) take the form
\begin{equation}\label{eqst}
\frac{1}{2}\partial_K g_{MN}
\partial_\mu X^M\partial_\nu X^N
\bAi^{\nu\mu}\sqrt{-\det\bA}
-\partial_\mu \left(
g_{KM}\partial_\nu X^M
\bAi^{\nu\mu}\sqrt{-\det\bA}\right)=0
\ . 
\end{equation}
Now let us restrict ourselves
to the case of  homogeneous
worldsheet fields  and also
presume that the 
 string is alligned along the
worldvolume of I-brane. Then
it is natural to take
\begin{equation}\label{gstr}
\tau=x^0 \ , \sigma=x^1 \ . 
\end{equation} 
For this ansatz the 
equation of motion 
(\ref{eqst}) for $X^1$ is trivially
satisfied since
\begin{equation}
\partial_\tau\left[
g_{11}\bAi^{\sigma \tau}\sqrt{-\det\bA}
\right]=0
\end{equation}
using the fact that for the time dependent
modes and for (\ref{gstr})
  the matrix $\bA$ takes the 
form
\begin{equation}
\bA_{\tau\tau}=g_{\tau\tau}+
g_{IJ}\partial_\tau X^I\partial_\tau X^J \ , 
\bA_{\tau\sigma}=\bA_{\sigma\tau}=0 \ ,
\bA_{\sigma\sigma}=g_{\sigma\sigma} \ . 
\end{equation}
Since  $g_{\tau\tau}=-1 \ ,
g_{\sigma\sigma}=1$ we 
immediatelly see that the fundamenal
string with vanishing
velocity can be localised
in any position labeled with
$(\bz,\by)$  
since  (\ref{stringfund})
does not  contain
any potential term and all 
terms as $g_{IJ}\dot{X}^I
\dot{X}^J$ vanish for 
$\dot{X}^I=0$.
This result is in agreement with 
the well known fact  that
the configuration of NS5-branes
and fundamental strings with the
paralell worldvolumes 
are supersymmetric. 

Let us now concetrate on the
configuration of coincident
NS5-branes where we have
\begin{equation}
H_1=1+\frac{\lambda_1} {|\bz|^2} \ ,
H_2=1+\frac{\lambda_2} {|\by|^2} \ ,
\end{equation}
and  
\begin{equation}
 \lambda_1=k_1l_s^2 \ ,
 \lambda_2=k_2l_s^2 \ .
\end{equation}
As in the previous section 
we use the manifest rotation
symmetries $SO(4)$ and introduce
 modes
\begin{equation}
z^6=R \cos \alpha \ ,
z^7=R \sin \alpha\ ,
\end{equation}
and
\begin{equation}
y^2=\tR \cos\beta \ ,
y^3=\tR \sin\beta \ .
\end{equation}
For this ansatz the string
action 
takes the form
\begin{equation}\label{sf}
s\equiv \frac{S}{V}
=-\tau_s\int d\tau
\sqrt{1-H_1(\dot{R}^2+R^2\dot{
\alpha}^2)-H_2(
\dot{S}^2+S^2\dot{\beta}^2)} \ . 
\end{equation}
In what follows we will be interested
in the study of the dynamics of
the fundamental strings 
in the near horizon region where
 $\frac{\lambda_1}
{R^2}\gg 1 \ , \frac{\lambda_2}{S^2}
\gg 1$.
In this region the action
simplifies as
\begin{equation}
s=-\int dt
\sqrt{1-\frac{\lambda_1}{R^2}
\dot{R}^2-\lambda_1\dot{\alpha}^2
-\frac{\lambda}{S^2}\dot{S}^2-
\lambda_2\dot{\beta}^2} \ . 
\end{equation}
Now using the transformations
\begin{equation}
R=e^{\frac{\phi_1}{\sqrt{\lambda_1}}} \ ,
S=e^{\frac{\phi_2}{\sqrt{\lambda_2}}}
\end{equation}
the action takes simple form
\begin{equation}
s=-\int dt
\sqrt{1-\dot{\phi}_1^2-
\dot{\phi}_2^2-
\lambda_1\dot{\alpha}^2-
\lambda_2\dot{\beta}^2}
\end{equation}
that explicitly 
demonstrates  the fact that F-string
and NS5-brane are supersymmetric
configurations since the dynamics
that follows from the action 
above corresponds to the motion
of free particle.


In this section we  studied
the dynamics of the fundamental string
probe in the I-brane background. We have
seen that this dynamics is trivial
since the probe F-string when it
is at the  rest, forms supersymmetric
configuration with the background 
I-brane. Even if the  dynamics is not very
interesting we mean that it was
useful to review it here since when
we consider I-brane in the type IIA
theory then the only stable probe
whose worldvolume fits the worldvolume
of I-brane is F-string. For
D2-brane one of its spatial dimensions
has to extend in the direction
transverse to the I-brane and hence
the  D2-brane DBI action 
 explicitly
depends on the worldvolume spatial coordinate.
Then  one can expect that the ansatz
with homogeneous fields  does not
solve the equation of motion that
arrise from the 
 DBI action for D2-brane 
 in I-brane background.

On the other hand there is a possibility
to probe I-brane with D0-brane whose
dynamics in I-brane background is
 governed by the action
\begin{equation}
S=-\frac{\tau_0}{g_1g_2}\int dt
\frac{1}{\sqrt{
H_1H_2}}\sqrt{1-H_1(\dot{R}^2+R^2\dot{\alpha}^2)
-H_2(\dot{S}^2+S^2\dot{\beta}^2)-
\dot{Y}^2} \ ,
\end{equation}
where $Y$ is the mode that describes
propagation of D0-brane along the
I-brane worldvolume. Now it is 
easy to see that the dynamics
of  D0-brane is equivalent to the
dynamics of the probe  D1-brane
with conserved electric flux. 
 This can be easily seen
from the fact that  $P_Y=\frac{\delta
L}{\delta \dot{Y}}$ is conserved
and it is related to the
conserved electric flux on the
worldvolume of D1-brane through T-duality.
 More preciselly,
if we compactify the theory on the
circle of radius R along the worldvolume
of I-brane in type IIB theory
we can see that now
the configuration of I-brane with
the probe of D1-brane is T-dual to
the I-brane on the dual circle and
D0-brane that moves with constant
momentum along this dual circle
in type IIA theory.
These arguments are based on the
fact that T-duality acts along worldvolumes
of the background of NS5-branes
and it is well known that under
this T-duality NS5-brane in type
IIA(IIB) theory is mapped to NS5-brane
in IIB(IIA) theory
\footnote{For review of this topic
and extensive list of references, see,
for example
\cite{Giveon:1998sr}.}.

Using this duality  it is now
easy to argue that D0-brane in
the near horizon region again ``feels''
an enhancement of the 
symmetry of the 
near horizon geometry 
\cite{Itzhaki:2005tu}
in the same way as it was shown 
in case of D1-brane in near horizon
region of I-brane background in 
\cite{Kluson:2005eb}.
\section{M5-branes overlapping in
a string}\label{fourth}
The solution studied in previous
section that consists two stacks
of orthogonal NS5-branes in type IIA
theory can be naturally uplifted
to the solution in M-theory that
has the form
\cite{Gauntlett:1996pb}
\begin{eqnarray}\label{M5}
ds^2=(H_1H_2)^{-1/3}
(-(dx^0)^2+(dx^1)^2)+H_1^{-1/3}
H_2^{2/3}\delta_{\alpha\beta}
dx^\alpha dx^\beta+\nonumber \\
H_1^{2/3}H_2^{-1/3}\delta_{pq}
dx^p dx^q+H_1^{2/3}H_2^{2/3}
(dx^{10})^2 \ , \nonumber \\
\end{eqnarray}
where $H_1,H_2$ have the same
form as in the previous section.
 There is
also nonzero three form $C$
however we will not need to know
its explicit from following reason.
We will be interested in the
dynamics of the probe M2-brane with
the worldvolume stretched along 
the line of intersection of M5-branes
 and also
along the direction $x^{10}$. Since
we also restrict ourselves to the
homogeneous modes it turns out that
the coupling of $C$ to the
M2-brane vanishes.

The background configuration (\ref{M5})
describes two stack of fivebranes
in the $(1,2,3,4,5)$ and in
$(1,6,7,8,9)$ directions
that overlap in a string in
$(1)$ direction. Note that
although the functions $H_1,H_2
$ depend on the relative transverse
directions, they are translationally
invariant in the overall 
transverse direction $x^{10}$. 

Now the natural probe of this
background is M2-brane
whose DBI part of the action takes
the form
\begin{equation}
S=-\tau_{M2}
\int d^3\xi
\sqrt{-\det g_{\mu\nu}} \ ,
g_{\mu\nu}=
g_{MN}\partial_\mu X^M
\partial_\nu X^N \ ,
\end{equation}
where $\tau_{M2}$ is M2-brane
tension and $\xi^\mu \ , \mu=0,1,2$
label the worldvolume of M2-brane.

If we fix the gauge as
$\xi^0=X^0=t \ , \xi^1=X^1 \ ,
\xi^2=X^{10}$ and  then restrict to 
the homogeneous modes 
we obtain the action $s\equiv \frac{S}{V_2}$
in the form
\begin{equation}
s=-\tau_{M2}
\int dt
\sqrt{-g_{11}g_{1010}
(g_{00}+g_{IJ}\partial_0X^I
\partial_0X^J)} \ , 
\end{equation}
where $V_2$ is the volume of
the spatial section of
M2-brane worldvolume and 
$X^I \ , I=2,\dots,9$ label the 
positions of
M2-brane in a  space transverse to 
its worldvolume.
Now if we insert the components of
the metric given in (\ref{M5})
and  introduce the
modes $R,S,\alpha,\beta$ in 
the same way  as in
the previous sections we obtain
the action
\begin{equation}\label{sact}
s=-\tau_{M2}
\int dt
\sqrt{1-H_1(\dot{R}^2+R^2\dot{\alpha}^2)
-H_2(\dot{S}^2+S^2\dot{\beta}^2)} \ .
\end{equation}
As we can expect this action has the
same form as the action for fundamental 
F-string in I-brane background 
studied in the previous section. This
follows from the fact that the
background (\ref{M5}) does not
depend on 
$x^{10}$. Moreover,  M2-brane is extended in
$x^{10}$  as well but we have
presumed that all worldvolume modes
depend on time only. Put differently, 
the  configuration of two
stacks of intersecting M5-branes and a probe
M2-brane with one spatial direction parallel with
the line of intersection of two
stacks of  M5-branes 
and the second one that is extended in $x^{10}$ direction
is supersymmetric \cite{Gauntlett:1996pb}. 
Then it follows that the action
(\ref{sact}) 
of such a probe in the near horizon
region after performing the same 
transformations as in the case of F-string
probe in I-brane background in type IIA
theory corresponds to the action
for free particle and hence its
dynamics is trivial. Since this
procedure is completely standard we
will not perform it here. 
\\
\\

{\bf Acknowledgement}

This work
 was supported in part by the Czech Ministry of
Education under Contract No. MSM
0021622409, by INFN, by the MIUR-COFIN
contract 2003-023852, by the EU
contracts MRTN-CT-2004-503369 and
MRTN-CT-2004-512194, by the INTAS
contract 03-516346 and by the NATO
grant PST.CLG.978785.


\end{document}